\documentclass[12pt,english]{article}
\newcommand{\di}{\rm d}

\begin{document}

\begin{center} 
{\Large\bf Thermodynamics } \\ [1mm]
{\Large\bf of the production processes }

\vspace{5mm}

{\large\it Vladimir Pokrovskii}\footnote{Corresponding author:  Vladimir Pokrovskii, vpok@comtv.ru} \\

\vspace{3mm}


Moscow State University of Economics, Statistics and Informatics, \\ Moscow, RUSSIA 119501

\vspace{5mm}

\end{center}

\centerline{Abstract}

The  process of creating goods and services, measured  by their value,  is considered  a process  of creating complexity.  This allows us to consider the production system as an open thermodynamic system, and to develop a simple heuristic model of the production process.    The model includes three production factors:  the index of complexity of  production equipment (physical capital $K$), human activity (labour $L$), and the substitutive capacity of equipment (substitutive work $P$).     The  thermodynamic approach  also needs in \ technological characteristics \ of production equipment,     \       such as labour \  requirement         $\overline{\lambda}$ and energy requirement   $\overline{\varepsilon}$, which indicate the amounts of labour and energy required to operate production equipment.     By applying thermodynamic principles, we can understand how labour can be replaced by capital and derive the  production function with   four  different formulations.  Two \ of them \  are  known  and  used by  researchers   for interpretation  the production \  phenomena;  \ the  thermodynamic  approach \ provides \ some \ foundation for  economic theory,  allowing us  to decompose unambiguously  the growth rate of output over  technological level  and the growth  rates of production factors.  Introducing  substitute work  as a factor of production   and technological  characteristics of capital expands our ability to model  and analyse production processes.

 \vspace{5mm}
\vspace{5mm}



$JEL$ - classification: C02,   E1	, O4

\vspace{5mm}


\newpage 

\section{Introduction}

 The relation  between the output of the production system and some quantities,   that are considered sources of value and  called production factors,  appears to be one of the main  pillars  of the theory of economic growth (the theory of production). The conventional theories (for example: Blanchard  and Fisher, 1989; Aghion and Howitt,  2009) attribute   output $Y$   to capital stock  $K$  and    labour input $L$, so that the relation  can   be written 
\begin{equation}
Y = Y (L, K). 
\end{equation} 
The relation (1) describes the effect of substitution. This means that two factors of production can substitute for one another. In other words, value creation is attributed not only to labour, which is considered sacred by all researchers, but also to installed equipment. Various specifications of functions are used by researchers, including some assumptions such as existence of the Solow residual or exogenous technical progress, if needed.

However, there is a puzzle regarding relation (1): how can labour be replaced by production equipment? We can hope that mechanisms of this replacement can be understood if we turn to thermodynamics. We are not seeking for analogies between economic and thermodynamic descriptions of the production  system, as some researchers have done, but rather we consider the production unit to be a collection of interacting particles that are organized in a complex structure. The production system ought to be considered as a complex thermodynamic system, which  is open for fluxes of substances and energy.

In the next section, we will refer to the concepts and principles of thermo-dynamics to describe the production process and  to investigate the relationship between output and consumed resources, that is  to develop a production function in various forms, one of which leads back to relation (1). A great number of works exist on the problem, including articles, which  use the thermodynamic terms, while interpretating the  production processes. These works motivated me to approach the production system as to an  open  thermo-dynamic system, the approach, which I believe could  place the theory of economic production  on a firmer thermodynamic footing. In any case, under-standing the quantitative relationship between outputs and resources can be useful in designing mathematical models of production.

 \newpage

\section{Heuristic model of the production process} 

 The mapping of the well-known production process,  the  simplest representation of which  is given in Figure 1,  with an open thermodynamic system   lies at the  heart  of our scheme.    In this section, we will discuss appropriate thermodynamic terms to describe the model.

\subsection{Production process as an open thermodynamic system} 

The purpose of production processes is to create useful things (food, clothing, buildings, transportation, sewage systems, household appliances, machines, etc.);    the  output of the production processes   $Y$    is  conventionally  interpreted as market exchange value  of all these products  estimated by  consumers.  However,  this  does not mean that it is a subjective quantity;  the objective content of output appears to be  complexity, which is a concept connected with entropy (Pokrovskii,  2020,  Chapter~8). The created complexity  $Y$  can be interpreted also as   the   diminishing of entropy   in the  human environment,  which allows us to state that the natural measure of value is energy (Beaudreau and Pokrovskii,  2010).  The  production of consumable goods  can be  interpreted    as a process of creating  useful complexity within the human environment;   production activity \  moves \ boundaries \ between the artificial \ and natural   environment.  

A core of the production system is a collection of equipment that is implemented as a set of tools for converting wild materials into useful forms. From a thermodynamic point of view, the production equipment in equilibrium is nothing more than a collection of interacting atoms and molecules.  The particles are  organised in the very complex structures,   which  can be seen as a thermodynamic non-equilibrium state  of the same system that is far from equilibrium. In order to characterize non-equilibrium states, as we know (Pokrovskii,  2020,  Chapter~2), apart from the usual variables associated with equilibrium states, we need to introduce additional characteristics, called internal variables or variables of complexity. To describe a production system's structure in detail, we need many internal variables; however, every non-equilibrium system  has a general characteristic --- the complexity index. Due to the genesis of production equipment  from the output of  production system, monetary estimate of the amount of equipment  $K$  appears to be a complexity
\linebreak

\setlength{\unitlength}{0.82mm}

\begin{center}
\begin{picture}(165,40)(0,80)



\put(10,140){\makebox(40,20){\footnotesize INPUT}}
\put(10,120){\framebox(40,19){\sf raw materials}}
\put(10,100){\framebox(40,19){\shortstack{{\it P}\\\vspace{0.2cm}\\ {\sf machine work}}}}
\put(10,80){\framebox(40,19){\shortstack{{\it L}\\ \vspace{0.2cm}\\{\sf labour efforts}}}}

 \put(50,130){\vector(1,0){15}}
 \put(50,110){\vector(1,0){15}}
 \put(50,90){\vector(1,0){15}}
 
\put(65,80){\framebox(40,65){\shortstack{{\it K}\\\vspace{0cm} 
\\{\sf production} \\\vspace{0cm}\\ {\sf equipment}\\\vspace{0.2cm}\\{\sf $\overline{\lambda}$ \ labour req.} \\\vspace{0cm}\\ {\sf $\overline {\varepsilon}$ \  energy req.}}}}
\put(68,93){\dashbox(34,16)}

 \put(105,130){\vector(1,0){15}}
 \put(105,110){\vector(1,0){15}}
 \put(105,90){\vector(1,0){15}}

\put(120,140){\makebox(40,20){\footnotesize  OUTPUT}}
\put(120,120){\framebox(40,19){\sf heat}}
\put(120,100){\framebox(40,19){\sf pollutants}}
\put(120,80){\framebox(40,19){\shortstack{{\it Y}\\\vspace{0.1cm}\\{\sf useful output}}}}

\end{picture}
\end{center}
\vspace{5mm}

\centerline{\bf Figure 1 \ Scheme of the production process} 

\vspace{2mm}

{\sl\small  \noindent   The  open thermodynamic system is being realized as a set of appliances for the matter transformations.   To produce a thing or service, apart from the production equipment  $K$, one  needs  raw materials (ores, water,   air, energy carriers and so on), worker efforts $L$ and machine work  $P$.  These  factors are  closely  related to technology    (the characteristics  $\overline{\lambda}$  and $\overline {\varepsilon}$ ), embodied in the  production equipment.   The useful output of the production process $Y$ is a monetary estimate of the collection  of useful things and services produced. Production processes are accompanied by heat and pollution emissions, but this is a separate issue;   the 'useful'  side of production is addressed in the article.}
\vspace*{2mm}
\hrule 

\vspace*{4mm}

\noindent 
 index in thermodynamic terms. An assembly of production equipment that is a specific embodiment of a technology can be viewed as a body of  an open thermodynamic system.

The production equipment is an essential part of modern enterprises. However, it remains inactive until the efforts of workers and the energy input from external sources are applied. Historically, classical political  economists (Smith,  1776/1976;  Marx, 1887/1954)    have chosen among  the three factors,  listed as input on  Figure~1,  the combined efforts of people --- labour  $L$ (efforts and work) --- as the sole source of value.  This was the perfect solution until machines started to dominate production processes. In order to solve the problems arising,  while  understanding production processes at the machine production stage, researchers added another factor to their description, coming to relation (1). 

The thermodynamic approach takes into account fluxes of energy coming to the production system from external sources.  The primary production energy is distributed throughout the production system, losing most of its amount  during various transfers. Only a small portion of the total energy $P$ consumed in production appears to reach production devices (Pokrovskii,  2003; 2018, Section 2.5). This small amount truly replaces labour in production processes and plays the role of an important production factor.  It  was a crucial decision to separate the substitutive work and introduce  it as a production  factor $P$ into the   theory.  Therefore, behind the substituting labour with capital,  situates  substituting labour with the work of production equipment.

To keep the  production system  in operation,  one needs in fluxes of external resources:  the fluxes of substances and the two forms of fluxes of energy,   one of them is  labour  $L$   ---  efforts and work of people;  the other ---  the work   done by production equipment  $P$,  using energy from outside sources such as wind, water or coal. Only  work in the forms of  human and machine work create useful complexity, which can be formalized as 
 \begin{equation}
Y  = Y(L, P).
\end{equation} 
The amounts of the used social resources,  $L$ and  $P$,  depends on the used technology, embodied in   the production equipment, measured  by its value~$K$.

\subsection{The main properties  of the production equipment} 

 The installed equipment determines  how much labour $L$ and substitutive work $P$ will be required for the production process before it is activated.  We assume that these factors of production  are  measured in energy units. 

To fix the  property of the capital to attract labour force and productive energy, we can, according to the previous speculations \  (Pokrovski, 1999, Chapter~4;  2018,  Chapter~5), introduce the main characteristics of the capital, which are labour and energy requirements, 
\begin{equation}
\lambda = \frac{\ d \, L}{\ d \, K}, \quad  \varepsilon = \frac{\ d \, P}{\ d \,  K}.		
\end{equation}
These quantities indicate the amounts of labour and productive energy required to operate  a unit of production equipment. 

It is convenient to deal with non-dimensional technological coefficients 
\begin{equation}
\overline \lambda (t) = \frac{K}{L}\,  \frac{\ d \,  L}{\ d \,  K},   \qquad 
\overline \varepsilon (t) = \frac{K}{P}\, \frac{\ d \, P}{\ d \,  K}.
\end{equation}
If the quantity is less than one, it means that labour-saving or energy-saving technologies have been used at that moment of time. In a stationary case, the quantities do not change with time.

The quantities (4) characterize the basic technological level of production equipment. They determine the amount of required production factors. If the non-dimensional technological coefficients are considered to be given characteristics, then the relations (4) appear to be equations that determine the required amounts of production factors
\begin{equation} 
 \frac {L}{L_0}=  \left(\frac{K}{K_0}\right)^{\overline \lambda}, \quad 
 \frac{P}{P_0} =  \left(\frac{K}{K_0}\right)^{ \overline \varepsilon}.
\end{equation}
The symbols with zeros represent the values of the variables at a selected starting point.

The  efficiency of the use of production factors is characterized by a technological index, which is a combination of dimensionless technological coefficients 
\begin{equation}
\alpha = \frac{1 - \overline{\lambda}}{\overline{\varepsilon} - \overline{\lambda}}.
\end{equation}

The technological coefficients and technological index are internal parameters of the production system. Variable $K$ represents the total amount (in monetary units)            of production equipment, while the technological coefficients $\overline{\lambda}$ and $\overline\varepsilon$ indicate its quality. 	 We assume that these three quantities are given as the global characteristics of production equipment.

\section{Specification of the production function} 

This section considers the output of the production process $Y$, as the market value of the product created at a given point in time, in the relation to its value $Y_{0}$ at an arbitrary initial point. This value can be expressed as a function of the production factors, $L$, $P$, and $K$, relative to their values at the beginning of the period: $L_{0}$, $K_{0}$, and $P_{0}$. It will be convenient to use the dimensionless ratios of these variables.

From a thermodynamic point of view, the true and only sources of value (as relatives of entropy) are labour expenses $L$ and the work of external energy sources $P$.      Human efforts are  applied directly, whereas substitutive work arises from the prior exploration of natural laws, the development of technological systems, and the manufacturing of production machinery, among other things.  The interpretation of the production process  allows us to view the output $Y$ as a function of two variables: the activity of workers --- labour $L$ --- and the work done by production equipment --- substitutive work $P$,  which depends, according to equations (5), on the amount  of installed equipment $K$, so that we can write the following function
\begin{equation}
Y = Y [L(K), P(K)].		
\end{equation}
As the  amounts of work,  the production factors  are  interchangeable and, in this sense, they are equivalent in all  respect.  The Smith-Marx theory of labour value is supplemented by the law of substitution, which says that  work of third-party forces of nature via production equipment replaces human efforts. The labour acts  as labour plus work of equipment in complex in the production of value.  

The general form of dependence of output on production factors can be determined by simple assumptions. Universality of production function requires that a proposed function can be applied not only to a specific case, but also to many different situations. For instance, the starting point can be arbitrary, but the shape of the function should not be influenced by this arbitrariness, which determines the unique power  form  of the function. Uniformity requires the  arguments of the function to be special combinations  of production factors.   These requirements lead (details can be found in the author's monograph (Pokrovskii,  2018, Chapter 6)  to the expression for the  production function in the form 
\begin{equation} 
Y = Y_0 \, \frac{L}{L_0}\left(\frac {L_0}{L} 
\frac{P}{P_0}\right)^\alpha, \quad 0 < \alpha < 1,
\end{equation}
The index $\alpha$ --- the technological index --- in equation (8) determines the degree of substitution of labour with productive energy. It was shown in (Pokrovskii,   1999, Chapter 4;    2018,  Chapter 6 ),  that the index $\alpha$ in equation (8) has the same form as it is defined  by the formula (6); the level of technology development is included in  the law of value production.  Note that the labour  $L$ and substitutive  work of external energy sources $P$ is measured in energy units.   Relation (8)  was confirmed in many details  in the author's previous works (Pokrovskii,  2003;    2018,  Chapter 6).

Relation (8) describes the effect of substitution labour efforts with machine work. Economists usually prefer to talk about substituting labour with capital (Aghion and Howitt,  2009). Relations (5) allow us to represent the production of value  (8) in terms of the variables $L$ and $K$. When we choose these variables, the resulting production function is as follows
 \begin{equation}
Y = Y_0 \, \frac{L}{L_0}\left(\frac {L_0}{L}\left[\frac{K}{K_0}\right]^{\overline\varepsilon} \, 
\right)^\alpha, \quad 0 < \alpha < 1. 
\end{equation}
The recorded relation is a specific form of function (1) and represents the law of labour substitution by capital. An increase in output can be achieved through increasing relative labour expenses, or the relative amount of capital.   The substitution coefficient depends on technological characteristics of production equipment.  An analysis of substitution mechanisms shows that workers' efforts to produce goods are substituted by the work of the production equipment, not by equipment itself. 

 There are various  empirical specifications of the dependence of output on the labour  $L$ and  physical capital $K$,  but no one coincides with thermodynamic results (9). Up to now the  existing empirical formulae for the calculation of output do not contain  the appropriate   technological characteristics of production equipment.  

Note that, due to the validity of equations (5), the functions (8) and (9) can be reduced to a simple relation 
  \begin{equation}
Y = Y_0 \, \frac{ K }{ K_0}. 
\end{equation}
The proportionality of  the output to the amount (in value) of capital is confirmed by observations (Blanchard and Fisher, 1989.  p. 4).  Existing deviations from this relationship may be due to different growth rates across industries (Pokrovskii 2018,  Chapter 9, section 9.3.2).

One of the relations (5) shows the amount of human effort for  the presence of production equipment (physical capital $K$), so that the relations (9) and (10) can be are represented in  the form
\begin{equation}
Y = Y_0 \,\left(\frac{ L}{L_0}\right)^{1/ \overline{\lambda}} .
\end{equation}
This relation gives us the reason to say that labour remains,  to use the words of Adam Smith, "the only universal, as well as the only accurate measure of value, or the only standard by which we can compare the values of various goods at all times and in all places".   The law of  the  value production appears to be a non-linear, generally speaking, function.

Expressions (8)-(11) represent four  forms of production function as a consequence of a thermodynamic approach.  The theory does not need to contain any arbitrary parameters  to be consistent with empirical observations. 

\section{The marginal productivity of capital}  

As can be seen from the production process description, the useful output     $Y$ is determined by three production factors, $L$, $P$ and $K$. Only two of them, $L$ and $P$, perform work; they  are active factors from thermodynamic point of view.  In contrast to this, capital  $K$ appears to be the means, which allows the active factors to perform  work, and  should be considered as a passive factor of production.

Four forms of production function give rise to four    different  interpretations of processes of value production . Economists usually prefer to interpret the situations  in terms of labour and capital  (Aghion and Howitt,  2009). Within the framework of this interpretation, the concept of the production power  of capital has been formed, and the idea of capital productivity has become ingrained in the public consciousness. This applies not only to physical capital, but also to financial capital in various forms. For example, if we own shares in a company, we receive dividends; if we deposit money in a bank, we earn interest. Shares and money represent capital in its broadest sense.

Thermodynamic interpretation allows us to find an expression for capital productivity.   To separate  the  individual  contribution of production factors to the production of value we refer to the so-called  marginal productivity of production factors. The active factors are characterized by the marginal productivities
\begin{eqnarray}
\beta = \frac{\partial Y}{\partial L}&=&Y_0 \, \frac{1 -\alpha}{L_0} \left(\frac {L_0}{L} \frac{P}{P_0}\right)^{\alpha},   \\  [2mm] 
 \gamma = \frac{\partial Y}{\partial P}&=&Y_0 \, \frac{\alpha}{P_0} \left(\frac {L_0}{L} \frac{P}{P_0}\right)^{\alpha-1}\nonumber. 		
\end{eqnarray}

Unlike factors $L$ and $P$, physical capital is not a source of value itself, but due to its role in production the `marginal productivity  of capital'   \,  can be formally calculated.    After differentiating relation (7) we find that.
\begin{equation}
\frac{d \, Y}{d \,K} = \beta \frac{\ d \,  L}{\ d \,  K} +  \gamma  \frac{\ d \, P}{\ d \,  K}		
\end{equation}

This expression shows that capital productivity is determined by marginal productivities of active production factors (12), and technological coefficients. One can see from equation (13) that the productive capacity of capital for the production  of  value is actually due to the productive capacity of labour and substitutive work.

\section{ Decomposition of the output growth rate}

Expressions  (8) --- (11)  demonstrate that, according to empirical data, at any moment the value of manufactured products $Y$, measured in monetary units of constant purchasing power, is determined by the technological characteristics of production equipment: energy requirement  $\overline{\varepsilon}$  and labour  requirement   $\overline{\lambda}$, also as  by  the aggregated amounts of social resources  (production factors) used:  physical capital $K$, direct  efforts of people $L$ (physical and mental, we cannot separate them) and the substitutive work of machines $P$  that is  obtained  with the help  of natural energy sources such as wind, water, coal and so on.  The  production factors are determined by the availability of social  resources and technological capabilities (the technological characteristics)  of the production equipment and are,   generally speaking, functions of time.

The influence of production factors and technological characteristics on {\it economic growth} is clearly demonstrated when calculating the growth rate of output $Y$, for which it is necessary to write down the full differential of output, referring to the relations (8) -- (11). For example, referring to the expression (8) and (11) allows one to write down the relations
\begin{equation}
\frac{1}{Y}\frac{\di Y}{\di t}= (1 - \alpha) \frac{1}{L}\frac{\di L}{\di t} +\alpha \frac{1}{P}\frac{\di P}{\di t}+ \ln \left(\frac {L_0}{L} \frac{P}{P_0}\right) \frac{\di \alpha}{\di t}.
\end{equation}
\begin{equation}
\frac{1}{Y}\frac{\di Y}{\di t}   =    \frac{1}{\overline{\lambda}} \left[\frac{1}{L}\frac{\di L}{\di t} -   \ln\left(\frac {L}{L_0} \right) \frac{1}{\overline{\lambda}} \frac{\di \overline{\lambda}}{\di t} \right].
\end{equation}

The growth rate of output is determined by the growth rate of production factors, while the 'structure' of the production system  remain unchanged,     and  by the technological characteristics of the equipment (the latter terms are on the right side of equations 14 and 15). With a constant value of the technological index, an increase in output is associated with an increase in the use of production factors with their constant characteristics, $P/L = const$, which is defined as {\it extensive economic growth}. The last term in formulae (14) and (15)  is directly related to structural and/or technological changes in the production system, that is, to the evolution of the production system itself. As the technological indexes  increase, the $P/L$ ratio also increases, and the case of {\it intensive economic growth} is realized.

To exclude the presence of zero values in the right-hand sides of ratios (14) and (15), which indicates that a change in technological characteristics with constant resource consumption ($L=L_o, P=P_o$) does not lead to any change in output, the last terms in these ratios can be written in a different form when using an expression that is valid in linear approximation by increment
$$
 \ln\left(\frac {L}{L_0} \right) \approx \frac{1}{L}  \Delta L.
$$
 The relations (14) and (15)  now take the form
\begin{equation}
\frac{1}{Y}\frac{\di Y}{\di t}= (1 - \alpha- \Delta \alpha) \frac{1}{L}\frac{\di L}{\di t} +(\alpha + \Delta \alpha) \frac{1}{P}\frac{\di P}{\di t}
\end{equation}
\begin{equation}
\frac{1}{Y}\frac{\di Y}{\di t}   =    \frac{1}{\overline{\lambda}} \left[1 -    \frac{1}{\overline{\lambda}}\, \Delta {\overline{\lambda}} \right]\frac{1}{L}\frac{\di L}{\di t}.
\end{equation}
The latter formula is convenient for analysing and evaluating the two policy options for increasing output: direct labour expenses  can be increased (additional labour is attracted) or technology and third-party energy consumption can be improved. The technological coefficient (labour  requirement)   $\overline{\lambda}$ decreases with increasing machine work, and with a decrease equal to the coefficient value, the effect of changing the technological characteristics of the system is compared with the effect of a direct increase in labour expenses.

For the actual decomposition of the output growth rate, time series on  output $Y$ and production factors are required. The decomposition of the output growth rate  is unambiguous and does not require additional information: there are no arbitrary parameters included in the theory.

\section{Concluding remarks }

No one seems to object to the fact that modern machine production is impossible without the consumption of external energy, but the question is whether energy consumption contributes to the production of value, which is the quantity  that appears in  inspection of exchange of products. In other words, the question is whether energy consumption can be included as an argument in the production function (there are different opinions on this matter), and if so, how.   The thermodynamic approach helps us to solve these fundamental questions on the basis of the statement that the energy consumed during the operation of production equipment replaces the efforts and work of people. This statement has allowed us to define and highlight the concept of substitute work as a production factor. The diversity of opinions about the role of energy in production has been influenced by the fact that energy is also consumed both as a final product (heating and lighting) and as an intermediate factor in production.  

Substitute work is a measure of all activities (discoveries, inventions, design, ...)   to replace the human efforts with machine work, and therefore the price of substitute work is not limited to the price of the small portion of primary energy carriers that can be attributed to this production factor. Substitute work $P$, as a production factor, has a special price. The amount of equipment used, which is necessary to support the substitute work $P$, should be estimated as $\mu K$,  $\mu$ being the coefficient of depreciation, so that the price of substitute work, as a production factor, is determined by the ratio 
\begin{equation}
p = \frac{\mu K}{P}
\end{equation}
This value is different from the price of energy carriers as ordinary intermediate or final products.

The fact that the expression for the output of a production system is written in four forms (8) - (11) helps to confirm  or refute the adequacy of the theory. Each of these relationships must be true if the theory is adequate. The main universal properties of technological equipment, measured by its value $K$, are defined by two technological characteristics: labour requirement $\overline{\lambda}$ and energy requirement $\overline{\varepsilon}$, which indicate the amounts of labour and energy required to operate the production equipment. The ratio (8)  did not meet the contradictions, when tested on the example of historical series on the US economy (Pokrovskii,   2003;    2018,  Chapter 6); the application of the theory is hampered by the difficulty of calculating the values of the substitutive  work. The  estimates for past years exist;  however, universal methods for calculating the substitutive  work have not been proposed.   One can be satisfied that form (10) is known and used by researchers for interpretation of the production phenomena.  The thermodynamic formula (9) appears to be a specific realisation of general function  (1) that is s a symbol of faith for economists.   It would be interesting to analyse the results of econometric studies on the relationship between output and production factors, and compare them with the results of econometric tests of the relationship (9). In all four cases, we should  find the same values of the technological coefficients if the proposed theory is accurate. 

According to the presented theory, technical progress, the main content of which  is  increase in consumption  of external energy,  that is the technical progress   is mainly  connected  with the creation and dissemination of devices that allow human labour to be replaced by machine work;  this  explains and describes the increase in labour productivity due to mechanization.   In addition to the processes of replacing human labour with machine work, there are other processes of substitution that should be considered. A person equipped with more advanced tools and using more suitable materials can produce more products with the same amount of labour. For example, if a worker uses a sharp iron axe instead of a dull stone axe, they can cut more wood per unit of time with the same amount of effort. In this case, the efficiency of the process of replacing human effort with machine work increases, which is described by the change  in the technological characteristics of physical capital                         $\overline{\lambda}$ and $\overline{\varepsilon}$. The dynamics of these quantities  also as dynamics of the  production factors was  described  earlier (Pokrovskii,   1999, Chapter 4;  2003;  2018,  Chapter 6;  2021).  

The thermodynamic approach helps us to solve one of the fundamental economic problem ---   the problem of origin of the wealth  in the epoch, when   different  machine and appliances  are widely used.  In this paper,  we have demonstrated  the law of the value production from the first principles; the production function does not contain any  extra suggestion and adjustable parameters.    This question is particularly relevant  for mathematical modelling  of  the  production processes:  one can hardly expect the sound results from the model,   when the initial relationship is not accurately represented.      The  thermodynamic  approach gives the solid  foundation for relationship  of  production theory.     The introduction of substitute work as a factor of production and the technological  features of capital    expands the ability to model and analyse  the economic processes.

\newpage

\section*{References}

\begin{itemize}

{\small

\item AGHION,  Ph. and HOWITT,  P.W. (2009). {\it The Economics of Growth}.  Cambridge, MA: MIT Press . 

\item BEAUDREAU, B.C.  and POKROVSKII, V.N.   (2010).   On the energy content of a money unit. {\it Physica A: Statistical Mechanics and its Applications},  389 (13):   2597 - 2606 .  

\item BLANCHARD O.J. and FISHER S.  (1989).   {\it Lectures on Macroeconomics}. Gambridge, MA: MIT Press. 

 \item MARX,  K.  (1887/1954).   {\it  Capital. Vol. 1.}  Moscow:   Progress Publishers.

\item  POKROVSKII, V.N.   (1999).   {\it Physical Principles in the Theory of Economic Growth}.   Aldershot:  Ashgate Publishing. 

 \item POKROVSKII, V.N.   (2003).   Energy in the theory of production.  {\it Energy}, 28(8): 769-788   

\item POKROVSKII, V.N.   (2018). {\it Econodynamics: The Theory of Social Production. The 3nd Ed.},   Dordrecht-Heidelberg-London-New York: Springer Nature. 

\item POKROVSKII, V.N.   (2020).  {\it Thermodynamics of Complex Systems: \  Principles and  applications}. Bristol, UK:  IOP Publishing. 

\item POKROVSKII, V.N.   (2021).   Social resources in the theory of economic growth.  {\it The Complex Systems},  No 3 (40): 33 - 44 . Available at Researchgate. 

 \item SMITH,  A. (1776/1976).  {\it An Inquiry into the Nature and Causes of the Wealth of Nations, in two volumes}. Oxford: Clarendon Press.
}
 \end{itemize}

\end{document}